\documentclass{jaa}
\usepackage{graphicx}
\usepackage{url}
\usepackage{xcolor}
\usepackage{float}

\begin{document}\sloppy

\title{\textit{AstroSat} Observation of 2016 Outburst of H~1743--322}


\author{Swadesh Chand\textsuperscript{1}, V. K. Agrawal\textsuperscript{2}, G. C. Dewangan\textsuperscript{3}, Prakash Tripathi\textsuperscript{3} and Parijat Thakur\textsuperscript{1,*}}
\affilOne{\textsuperscript{1}Department of Pure and Applied Physics, Guru Ghasidas Vishwavidyalaya (A Central University), Bilaspur (C. G.)-495009, India\\}
\affilTwo{\textsuperscript{2}Space Astronomy Group, ISITE Campus, ISRO Satellite Centre, Bangalore-560037, India\\}
\affilThree{\textsuperscript{3}Inter-University Centre for Astronomy and Astrophysics, Post Bag 4, Ganeshkhind, Pune-411007, India}


\onecolumn{

\maketitle

\corres{parijat@associates.iucaa.in; parijatthakur@yahoo.com}

\msinfo{...}{...}

\begin{abstract}
We present the detection of  type C quasi-periodic oscillation (QPO) along with upper harmonic at respective frequencies of $\sim0.6$ Hz and $\sim1.2$ Hz in the single AstroSat observation taken during the 2016 outburst of the low-mass black hole X-ray binary H~1743--322. These frequencies are found to be shifted by $\sim0.4$ Hz for the QPO and $\sim0.8$ Hz for the upper harmonic with respect to that found in the simultaneous \textit{XMM-Newton} and \textit{NuSTAR} observation taken five days later than the AstroSat observation, indicating a certain geometrical change in the system. However, the centroid frequency of the QPO and the upper harmonic do not change with energy, indicating the energy-independent nature. The decreasing trend in the fractional rms of the QPO with energy is consistent with the previous results for this source in the low/hard state. The value of the photon index ($\Gamma\sim1.67$) also indicates that the source was in the low/hard state during this particular observation. In addition, similar to the \textit{XMM-Newton} observations during the same outburst, we find a hard lag of $\sim21$ ms in the frequency range of $\sim1-5$ Hz. The log-linear trend between the averaged time lag and energy indicates the propagation of fluctuations in the mass accretion rate from outer part of the accretion disk to the inner hot regions.
\end{abstract}


}


\doinum{12.3456/s78910-011-012-3}
\artcitid{\#\#\#\#}
\volnum{000}
\year{0000}
\pgrange{1--}
\setcounter{page}{1}
\lp{16}

\section{Introduction}
A majority of black hole X-ray binaries (BHXRBs) exhibits transient nature and shows occasional outbursts due to sudden change in the mass accretion rate while spending most of the time in quiescence. The source luminosity may increase up to several orders of magnitude during such outbursts (Tanaka \& Shibazaki 1996; Shidatsu {\em et al.} 2014; Plant {\em et al.} 2015). In the course of a usual outburst, the black hole transients (BHTs) evolve through the low/hard state (LHS) to the high/soft state (HSS) via two intermediate states, viz. the hard and soft intermediate states (HIMS and SIMS; Belloni {\em et al.} 2005; Belloni 2010). These states are attributed to certain spectral and timing characteristics, which can be distinguished through the hardness intensity diagram (HID;  Belloni {\em et al.} 2005; Homan \& Belloni 2005; Gierli{\'n}ski \& Newton 2006; Fender {\em et al.} 2009; Belloni 2010). The X-ray spectrum in the LHS is dominated by the Comptonized emission with a powerlaw index $<2$ and cutoff energy $\sim100$ keV, and the source is associated with strong variability ($\sim 30\%$). On the other hand, the thermal emission from the optically thick and geometrically thin accretion disk dominates the HSS, where the photon index can extend up to $2.5$ with a few percent of variability. 

Low-frequency quasi-periodic oscillations (LFQPOs), ranging from $0.05-30$ Hz, are often observed in the BHTs. The exact origin of these LFQPOs are still not clear. However, LFQPOs are categorized into three types: type A, B and C. Among all the three types, the type C QPOs are very common, and appear as strong and narrow variable peaks with strong fractional root-mean-squared (rms) variability ($\sim3-16\%$) in both the LHS and HIMS. Type B QPOs are typically observed in the intermediate states having rms up to $\sim4\%$, whereas type A QPOs appear as broad peaks with a few percent of rms (Casella, Belloni \& Stella 2005; Motta {\em et al.} 2011; Alam {\em et al.} 2014).

BHTs show variability over the timescale of few seconds to days and the variability relies upon the change in source flux and energy due to different ongoing physical processes. One of the most prominent approach to investigate variability observed in BHTs is the study of time lag between the different energy bands. The measured time lags can be either positive or negative. The positive or hard lag implies that hard photons are delayed relative to the soft ones, and is supposed to be caused by the propagation of mass accretion rate fluctuation in the accretion disk (Page {\em et al.} 1981; Miyamoto {\em et al.} 1988; Lyubarskii 1997; Nowak {\em et al.} 1999a, 1999b; Grinberg {\em et al.} 2014; Ar{\' e}valo \& Uttley 2006; De Marco {\em et al.} 2013). On the other hand, the coronal X-rays get reflected from the accretion disk and give rise to a time delay between the primary X-ray continuum and the reprocessed emission from the inner accretion disk close to the central source. This time delay is known as the soft lag or the reverberation lag (De Marco {\em et al.} 2013; Kara {\em et al.} 2014; De Marco \& Ponti 2016; De Marco {\em et al.} 2017; Kara {\em et al.} 2019).

H~1743--322, a low-mass black hole transient source discovered with \textit{Ariel-V} in 1977, shows frequent outburst over the time scale of $\sim200$ days (Kaluzienski \& Holt 1977; Shidatsu {\em et al.} 2012, 2014). The first brightest outburst of this source took place in 2003, and was observed by \textit{INTEGRAL} (Revnivtsev et al. 2003). \textit{RXTE} also observed the same outburst, leading to the detection of a pair of high-frequency QPOs (HFQPOs; Homan {\em et al.} 2005; Remillard {\em et al.} 2006). Steiner {\em et al.} (2012) estimated the spin of the black hole to be $0.2\pm0.3$ with \textit{RXTE} observation during the 2003 outburst, and used the radio observation by very Large Array (\textit{VLA}) telescope of the same outburst to estimate the source distance and inclination angle  to be $8.5\pm0.8$ kpc and $75^\circ\pm3^\circ$, respectively. Moreover, Sriram {\em et al.} (2009) detected QPO in the steep powerlaw state of the source during the 2003 outburst, observed by \textit{RXTE}. H~1743--322 has also gone through several outbursts until 2008. This outburst in 2008 was found to be a failed one as the source could not reach the HSS over the entire outburst cycle due to abrupt change in the mass accretion rate (Capitanio {\em et al.} 2009). A few more outbursts of the source were detected between 2009 and 2013 by several space observatories. Altamirano \& Strohmayer (2012) discovered mHz QPO using the \textit{RXTE} and Chandra observations of 2010 and 2011 outbursts. Using the \textit{RXTE} observations of the 2010 and 2011 outbursts, Molla {\em et al.} (2017) also detected the presense of QPO at $\sim1$ Hz, and estimated the mass of the black hole to be $11.21^{+1.65}_{-1.96}$ M$_\odot$. Apart from the above, another outburst of the source in 2014 was observed by \textit{XMM-Newton}, \textit{NuSTAR} and \textit{Swift}/XRT. From the hardness intensity diagram (HID) derived from the \textit{Swift}/XRT monitoring, Stiele \& Yu (2016) found that the 2014 outburst was a failed one as the source stayed in the LHS throughout the full outburst. They also detected QPO ($\sim0.25$ Hz) along with upper harmonic ($\sim0.51$ Hz) in the commensurate ratio of $1:2$, and an iron emission line around $\sim6.7$ keV in the \textit{XMM-Newton} observation. The accretion disk, estimated from the relativistic reflection model was found to be truncated during this outburst (Ingram {\em et al.} 2017). A soft X-ray lag of $\sim60$ ms, arising most probably due to thermal reverberation, was also detected by De Marco \& Ponti (2016) in the \textit{XMM-Newton} observations of both the 2014 and 2008 failed outbursts. Furthermore, Chand {\em et al.} (2020) analyzed the two consecutive simultaneous \textit{XMM-Newton} and \textit{NuSTAR} observations of the 2016 outburst of H~1743--322. Unlike the 2008 and 2014 failed outbursts, they found the outburst in the 2016 to be a successful one using the HID derived with the \textit{Swift}/XRT monitoring of the same outburst. They found the presence of the QPO and its upper harmonic at $\sim1$ Hz and $\sim2$ Hz, respectively. Moreover, the centroid frequencies of the QPO and its upper harmonic were found to be drifting to higher frequencies over the two epochs. They also detected a fluorescent iron emission line at $\sim6.5$ keV and found the accretion disk likely to be truncated. A hard X-ray lag of $0.40\pm0.15$ s and $0.32\pm0.07$ s in the two respective epochs were also found by them. It is worth mentioning here that \textit{AstroSat} also observed the 2016 outburst of H~1743--322 five days earlier than the above mentioned simultaneous \textit{XMM-Newton} and \textit{NuSTAR} observations. Hence, it would be interesting to compare the results from the \textit{AsroSat} with those obtained from the above mentioned simultaneous \textit{XMM-Newton} and \textit{NuSTAR} observations for the 2016 outburst. Broadband coverage of \textit{AstroSat} can also shed light on the timing properties of the source at higher energy ($>30$ keV), which was not possible with earlier space observatories. In this paper, we carry out the timing and broadband ($0.7-80$ keV) spectral study of H~1743--322 using \textit{AstroSat} observation. 

The paper is organized as follows. In Section 2, we describe the observation and data reduction. Section 3 presents the analysis and results. Finally, the discussion and concluding remarks are given in Section 4.


\begin{figure}[H]
\includegraphics[width=0.9\columnwidth]{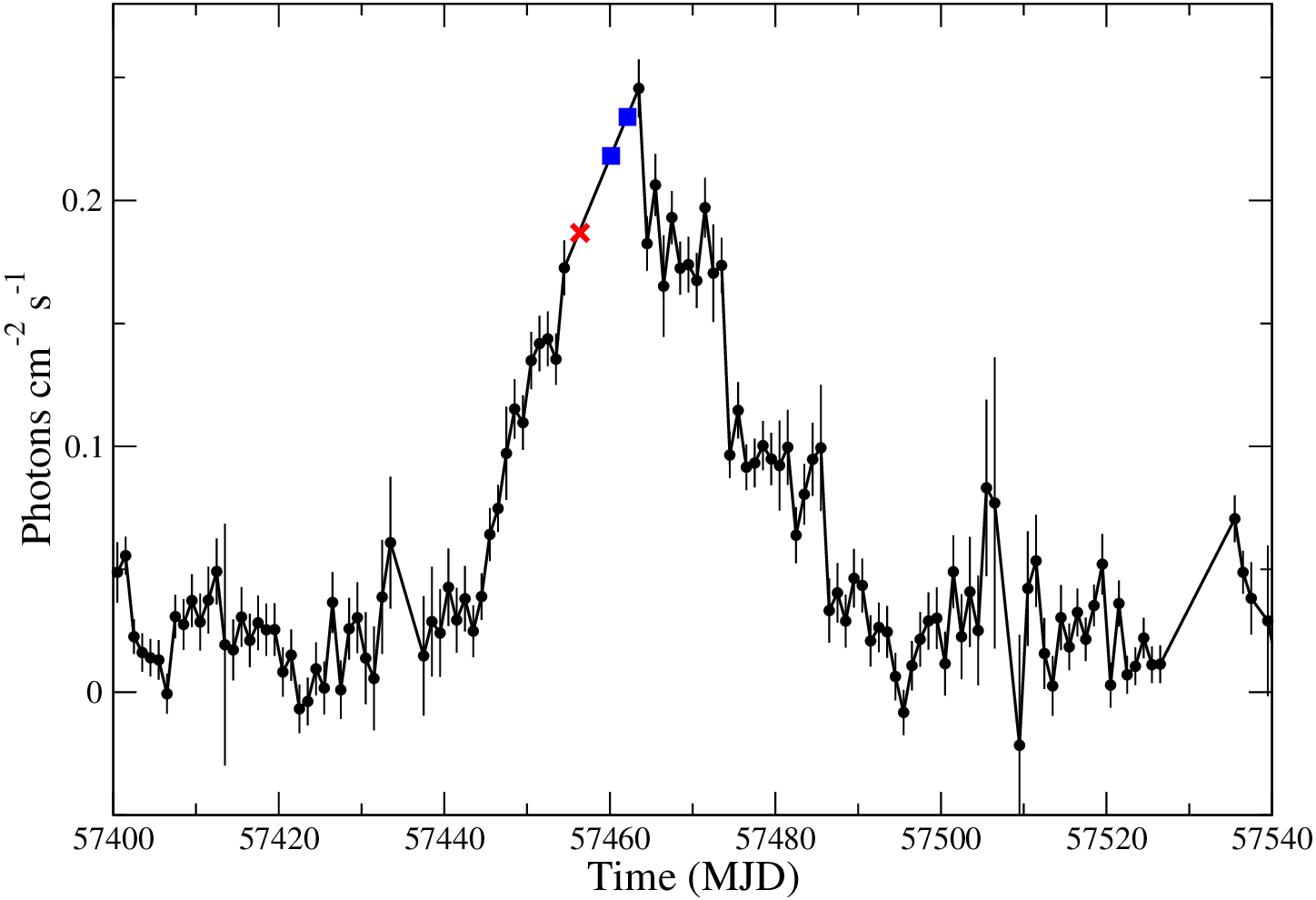}
\caption{MAXI lightcurve during the 2016 outburst of H~1743--322. The two blue sqaures indicate the positions of the two XMM-Newton observations during the same outburst, whereas the AstroSat observation is depicted with the red cross sign.}\label{figOne}
\end{figure}

\section{Observation and Data Reduction}
We used the single \textit{AstroSat} target of opportunity observation (Obs Id: T01$\_$045T01$\_$9000000364) of H~1743--322 taken during its 2016 outburst phase for an effective LAXPC exposure time of $\sim12.6$ ks. This observation was taken five days earlier than the simultaneous \textit{XMM-Newton} and \textit{NuSTAR} observations of the same outburst. Figure 1 shows the \textit{MAXI} lightcurve indicating the position of the \textit{AstroSat} observation along with the simultaneous \textit{XMM-Newton} and \textit{NuSTAR} observations during the mentioned outburst period of the source.

The level2 data of Soft X-Ray Telescope (SXT; Singh {\em et al.} 2016, 2017) observations were downloaded from the ISSDC website\footnote {\url {https://astrobrowse.issdc.gov.in/astro_archive/archive/Home.jsp}}. Using the SXT event merger tool\footnote{\url {https://www.tifr.res.in/~astrosat_sxt/dataanalysis.html}}, we obtained an exposure-corrected merged single cleaned event file. To extract the source spectrum, we used the standard available HEASoft tool XSELECT V2.6d and a circular region of 15$^{'}$ from the merged clean event file. For the response matrix file (RMF) and background spectrum, we used the SXT POC team provided ``sxt$\_$pc$\_$mat$\_$g0to12.rmf" and the blank sky observation ``SkyBkg$\_$comb$\_$EL3p5$\_$Cl$\_$Rd16p0$\_$v01.pha", respectively. By employing the on-axis ARF provided by the SXT POC team, the SXT off-axis auxiliary response file (ARF) was generated with the sxtARFModule tool\footnote{\url{https://www.tifr.res.in/~astrosat_sxt/dataanalysis.html}}, which is suitable for the location of the source on the CCD. The SXT spectrum in the $0.7-6$ keV band was used for sepctral analysis. Since the current version of the SXTPIPELINE does not correct for the gain shifts, SXT team recommends gain correction by keeping the slope fixed at unity and the offset to be variable. During the fitting of the SXT spectral data, this same method was applied to modify the gain of the response file.

We obtained the level2 data of Large Area X-Ray Proportional Counter (LAXPC; Yadav {\em et al.} 2016a, 2016b; Agrawal {\em et al.} 2017; Antia {\em et al.} 2017) using the LAXPC software (Laxpcsoft\footnote{\url{http://astrosat-ssc.iucaa.in/?q=laxpcData}}). The standard task within Laxpcsoft\footnote{\url{https://www.tifr.res.in/~astrosat_laxpc/LaxpcSoft.html}} were employed to derive the energy spectrum and light curve. For spectral analysis, we used only the LAXPC20 data due to its low background.


\begin{figure}[H]
\includegraphics[width=0.9\columnwidth]{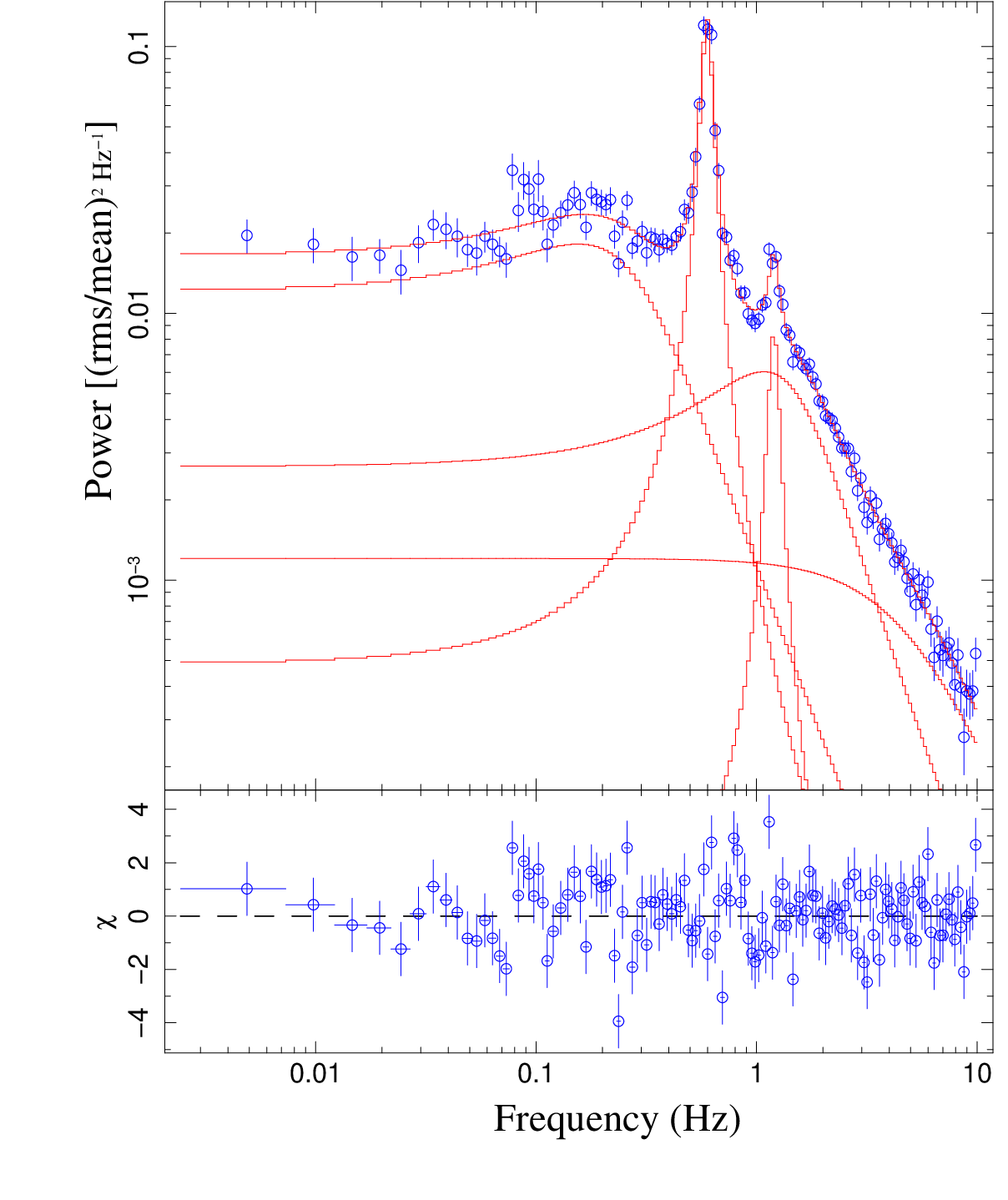}
\caption{Power density spectra in the $3-15$ keV band fitted with five Lorentzians.}\label{figOne}
\end{figure}

\begin{figure}[H]
\includegraphics[width=0.9\columnwidth]{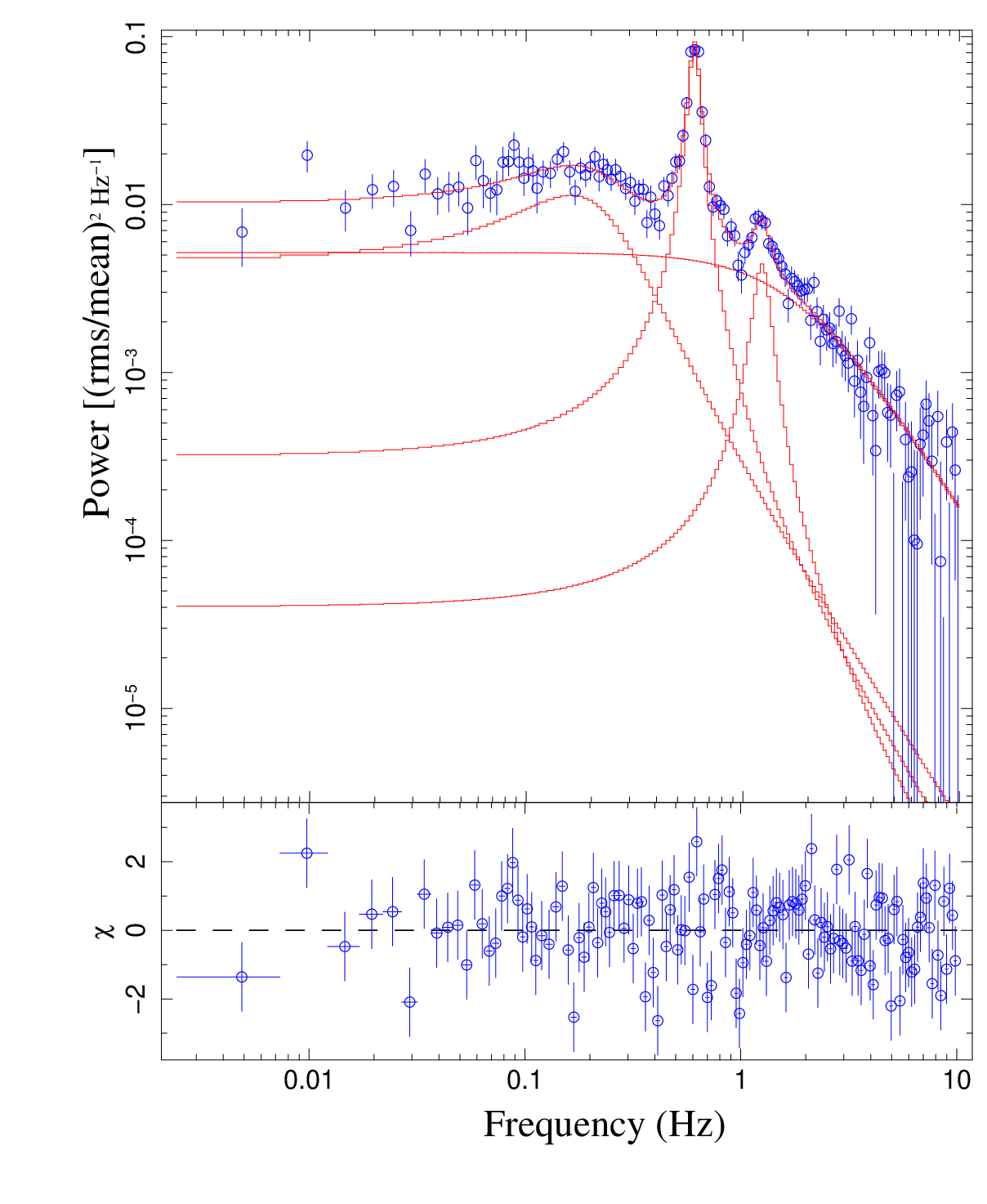}
\caption{Power density spectra in the $15-30$ keV band fitted with four Lorentzians.}\label{figOne}
\end{figure}

\begin{figure}[H]
\includegraphics[width=0.9\columnwidth]{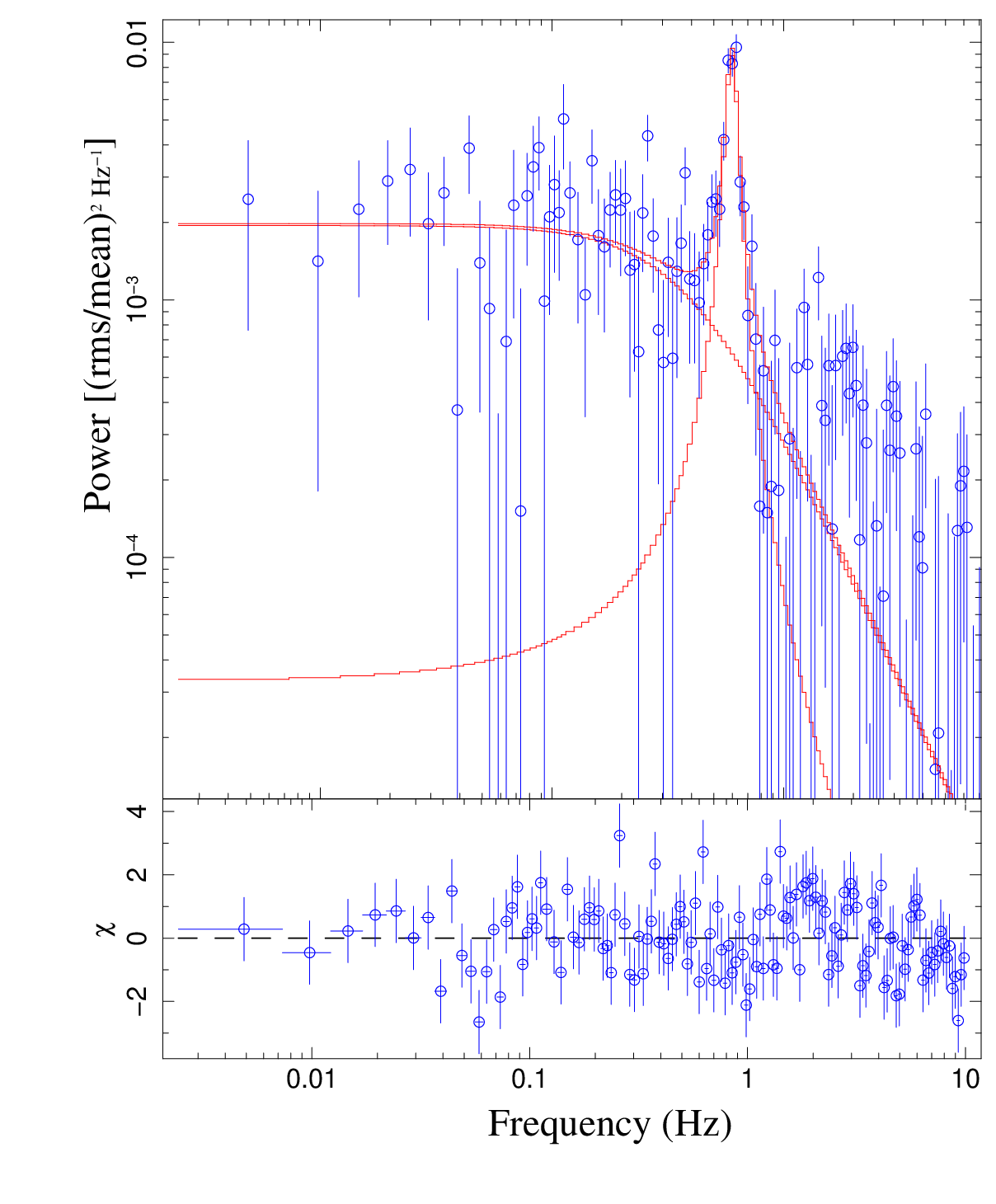}
\caption{Power density spectra in the $30-80$ keV band fitted with two Lorentzians.}\label{figOne}
\end{figure}

\section{Analysis and Results}

\subsection{Timing Analysis}

We have used Interactive Spectral Interpretation System (ISIS, V.1.6.2--40; Houck \& Denicola 2000) for the timing analysis and quoted the errors in 90\% confidence level. For the extraction of light curves, both the LAXPC10 and LAXPC20 detectors were used. Power density spectra (PDSs) from the lightcurves of 0.05 s bin were derived with ``POWSPEC" task within FTOOLS in the three bands: $3-15$, $15-30$ and $30-80$ keV. The PDSs were normalized as per Leahy {\em et al.} (1983) after the subtraction of contribution from the Poisson noise, and then the variability power was converted to the square fractional rms (Belloni \& Hasinger 1990). The PDSs derived in all the three bands shows the presence of quasi-periodic oscillation (QPO) at 
$\sim0.6$ Hz. Also the presence of an upper harmonic at $\sim1.2$ Hz is found in the $3-15$ and $15-30$ keV bands. However, it was absent in the $30-80$ keV band, which may be due to poor signal to noise ratio in the data at such high energy. Figure 2 shows the PDS derived in the $3-15$ keV band fitted with five Lorenztians required for the QPO, upper harmonic and three band limited noise (BLN) components. In addition to the QPO and its upper harmonic, two additional Lorentzains were employed to describe the two BLN components for the PDS derived in the $15-30$ keV band (see Figure 3). However, only two Lorentzians, required for the QPO and a single BLN component, were employed to describe the PDS in the $30-80$ keV band (see Figure 4). The best fit model parameters are listed in Table 1. It can be seen from Table 1 that the QPO and its upper harmonic are present in the $1:2$ ratio, and their centroid frequencies do not change with energy. For all the three bands, the quality-factor (Q$=\nu_{centroid}$/FWHM) of the QPOs are found to be similar within the errors. The same is also true for the upper harmonic in the $3-15$ and $15-30$ keV bands. The fractional rms amplitude of the QPO in the $15-30$ keV band shows a slight decrease than that found in the $3-15$ keV. However, it decreases marginally in the $30-80$ keV band. The fractional rms of the upper harmonic remains similar in the $3-15$ and $15-30$ keV bands.
Moreover, the significance of the QPOs are $\sim30.6\sigma$ (for $3-15$ keV), $\sim26.5\sigma$ (for $15-30$ keV) and $\sim11.7\sigma$ (for $30-80$ keV), which suggests that it is decreasing with increase in energy. In a similar way, the significance of the upper harmonic is $\sim8.7\sigma$ and $\sim5.2\sigma$ in the $3-15$ keV and $15-30$ keV bands, respectively.

\vspace{4em}

\begin{figure}[H]
\includegraphics[width=0.9\columnwidth]{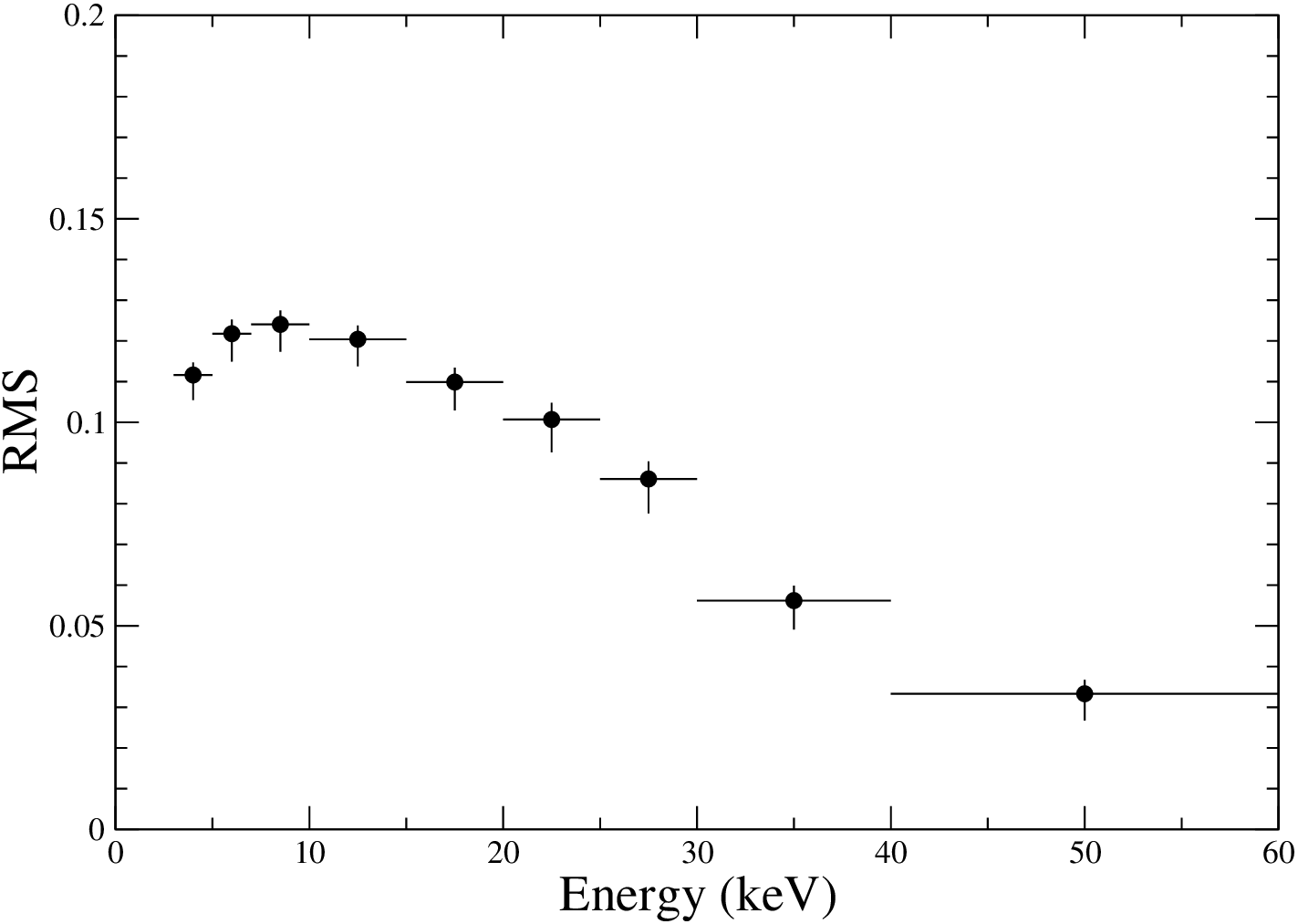}
\caption{Evolution of the fractional rms of QPO with energy.}\label{figOne}
\end{figure}

\begin{figure}[H]
\includegraphics[width=.9\columnwidth]{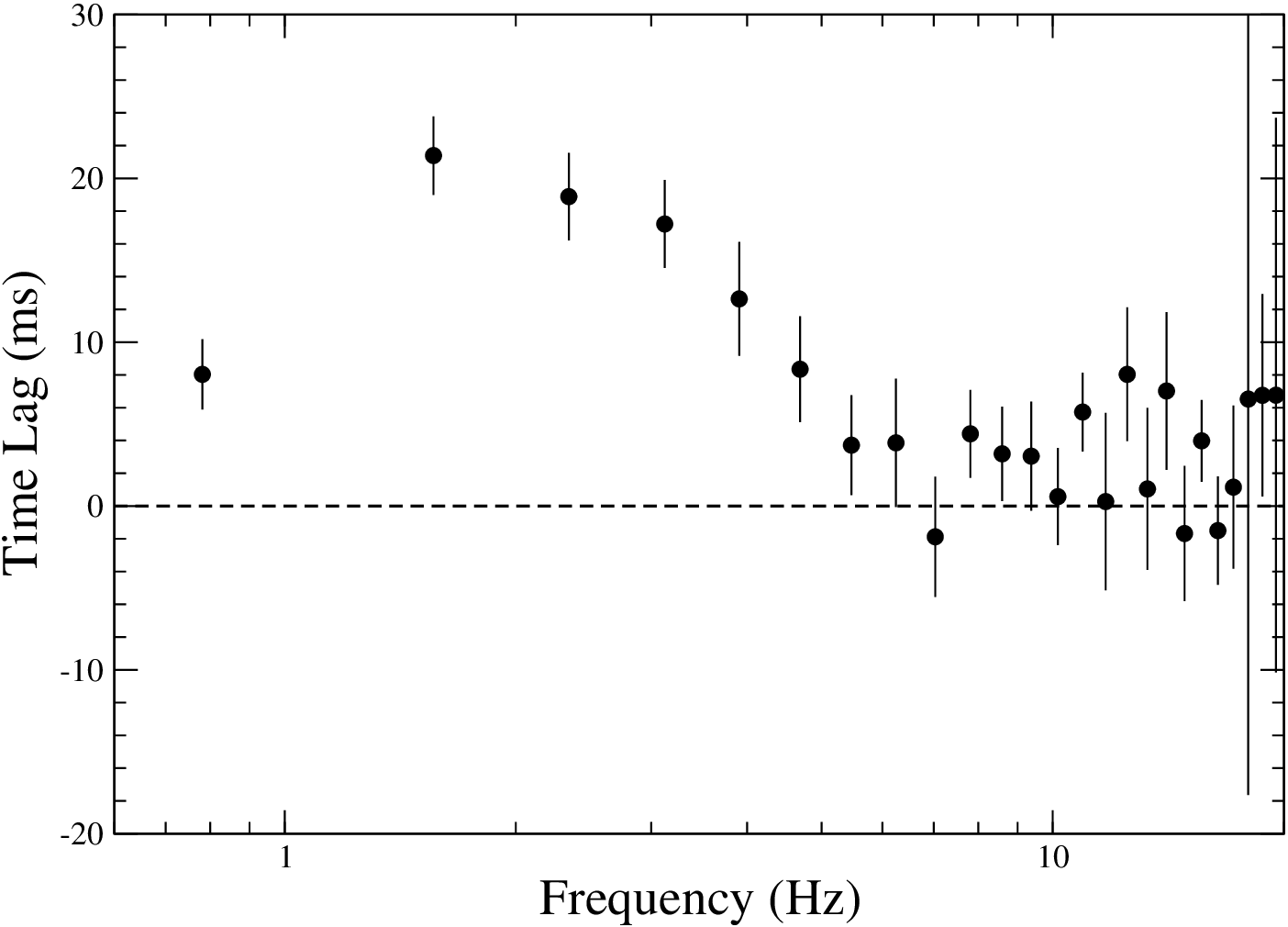}
\caption{Frequency dependent time lag between $3-10$ keV and $20-40$ keV band.}\label{figOne}
\end{figure}

To study the behavior of the fractional rms of the QPO as a function of energy, we extracted PDSs in different energy bands. These PDSs were fitted with five Lorentzians, where all the parameters except the normalization were kept fixed at the values obtained for the PDS in the $3-15$ keV band. Figure 5 depicts that the fractional rms of the QPO shows a decreasing nature with the increase energy.
To estimate the time lag as a function of frequency, we used the LAXPC subroutine `laxpc\_find\_freqlag', which employs the method as described in Vaughan \& Nowak (1997) and Nowak et al. (1999a). In this method, time lag between two different time series $s(t)$ and $h(t)$ is calculated as $\tau(f)=\phi(f)/2\pi f$, where $\phi(f)=arg[C(f)]$, is the phase of the average cross power spectrum $C(f)$. The cross power spectrum $C(f)$ is defined as $C(f)=S^*(f)H(f)$, where $S(f)$ and $H(f)$ represent the discrete Fourier transforms of the mentioned time series $s(t)$ and $h(t)$, respectively. 
As shown in Figure 6, we found a positive lag of $21.4\pm2.4$ ms in the frequency range $\sim1-5$ Hz between the soft photons in the $3-10$ keV band and the hard photons in the $20-40$ keV band. This indicates that the high energy photons in the $20-40$ keV band lag the soft ones in the $3-10$ keV band. The coherence between the above mentioned energy bands were found to be nearly equal in the frequency range of $1-5$ Hz. This frequency range was used to estimate the averaged time lag as function of energy by considering the same energy bands as used for the derivation of the QPO fractional rms. The averaged time lag estimated in the frequency range $1-5$ Hz shows an increasing trend with energy in a log-linear manner (see Figure 7).
\vspace{1em}

\begin{figure}[H]
\includegraphics[width=.9\columnwidth]{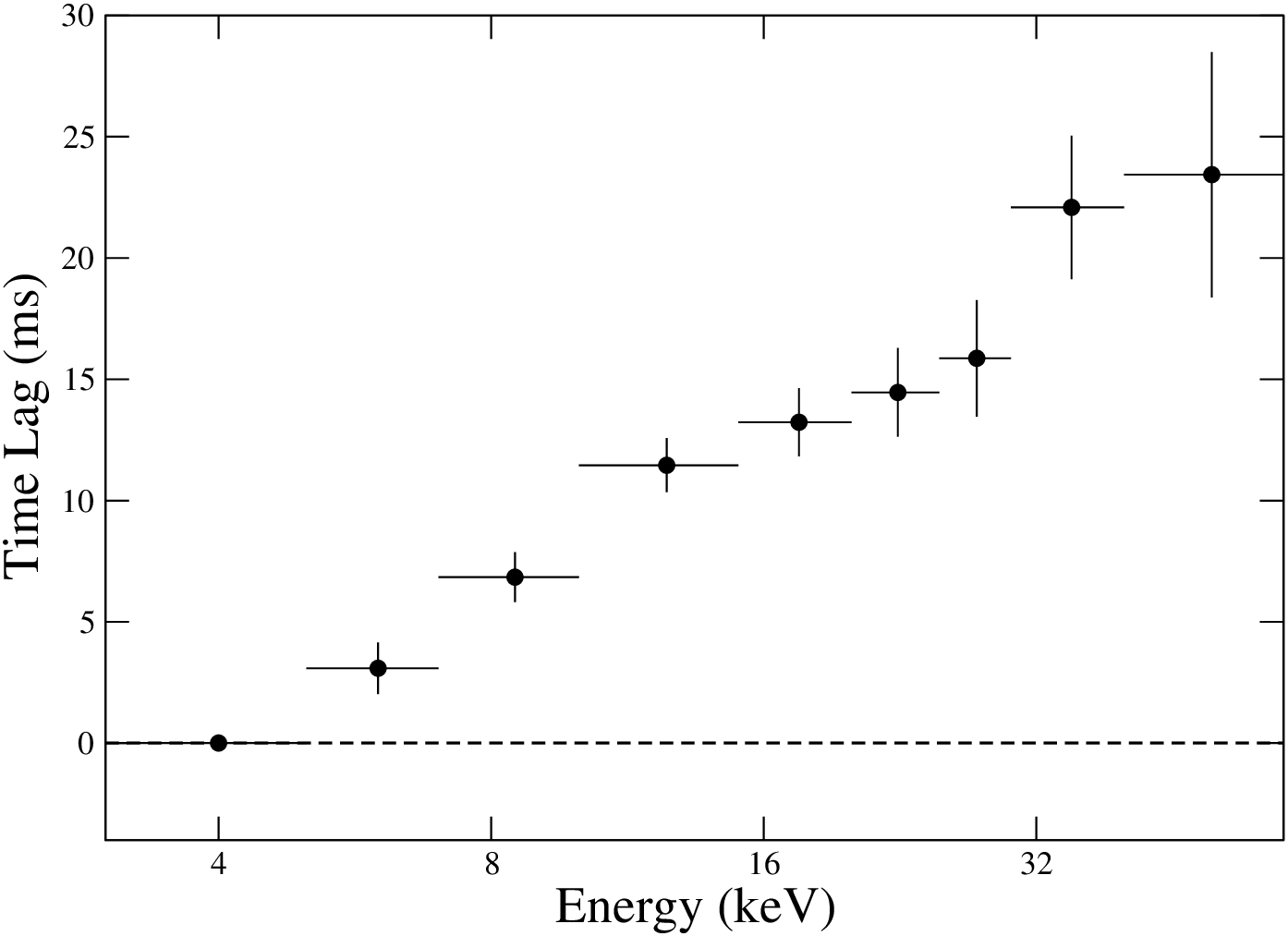}
\caption{Evolution of the averaged time lag with energy. The reference band is always the 3-5 keV band, corresponding to zero time lag point.}\label{figOne}
\end{figure}

\subsection{Broadband Spectral Analysis}

Time averaged broadband X-Ray spectral analysis was performed by fitting the SXT ($0.7-6$ keV) and LAXPC20 ($3-80$ keV) spectral data simultaneously within  ISIS (V.1.6.2--40; Houck \& Denicola 2000). All the error bars are quoted in 90\% confidence level unless otherwise specified. To account for the calibration uncertainty, we added a $3\%$ systematic uncertainty to SXT and LAXPC20 data. We grouped the LAXPC20 spectral data to a minimum signal to noise (S/N) ratio of 5 and a minimum of 1 channel per bin. For the SXT data, we used a minimum S/N of 10 and a minimum of 3 channels per bin. We fitted the spectral data from both the instruments simultaneously with multi-color disk blackbody ({\tt diskbb}; Mitsuda {\em et al.} 1984) modified by the Galactic absorption model {\tt tbabs} with the abundances of Wilms {\em et al.} (2000) and the cross section given by Verner {\em et al.} (1996). A multiplicative constant model was also used to take care of the relative normalization between the two instruments. The constant factor was fixed at 1 for the LAXPC20 data and kept free to vary for the SXT data. To account for the thermal compotonization from the hot corona, we used the {\tt nthcomp} model (Zdziarski {\em et al.} 1996; {\. Z}ycki {\em et al.} 1999) and tied the seed photon temperature (KT$_{bb}$) with the inner disk temperature (KT$_{in}$) from the {\tt diskbb} model. Hence, the model {\tt constant*tbabs(diskbb+nthcomp)} provides a best fit with $\chi^2$/dof $=204.5/248$. The best fit model to the SXT and LAXPC20 spectral data is shown in Figure 8, whereas all the best fit model parameters from the broadband X-Ray spectral fitting are listed in Table 2. Unlike the \textit{XMM-Newton} and \textit{NuSTAR} observations of the same outburst of this source (see Chand {\em et al.} 2020), we could not detect the presence of iron emission line  from the five days earlier observation by \textit{AstroSat}. As seen from Table 2, the hydrogen column density is found to be 1.94$\pm0.07$. The value of the photon index ($\Gamma \sim1.67$) indicates that the source was in the LHS during this observation. The electron plasma temperature (KT$_{e}$) is found to be $>56.9$ keV. Moreover, we found the inner disk temperature (KT$_{in}$) to be high $\sim1.2$ keV.  

\begin{figure}[H]
\includegraphics[width=.9\columnwidth]{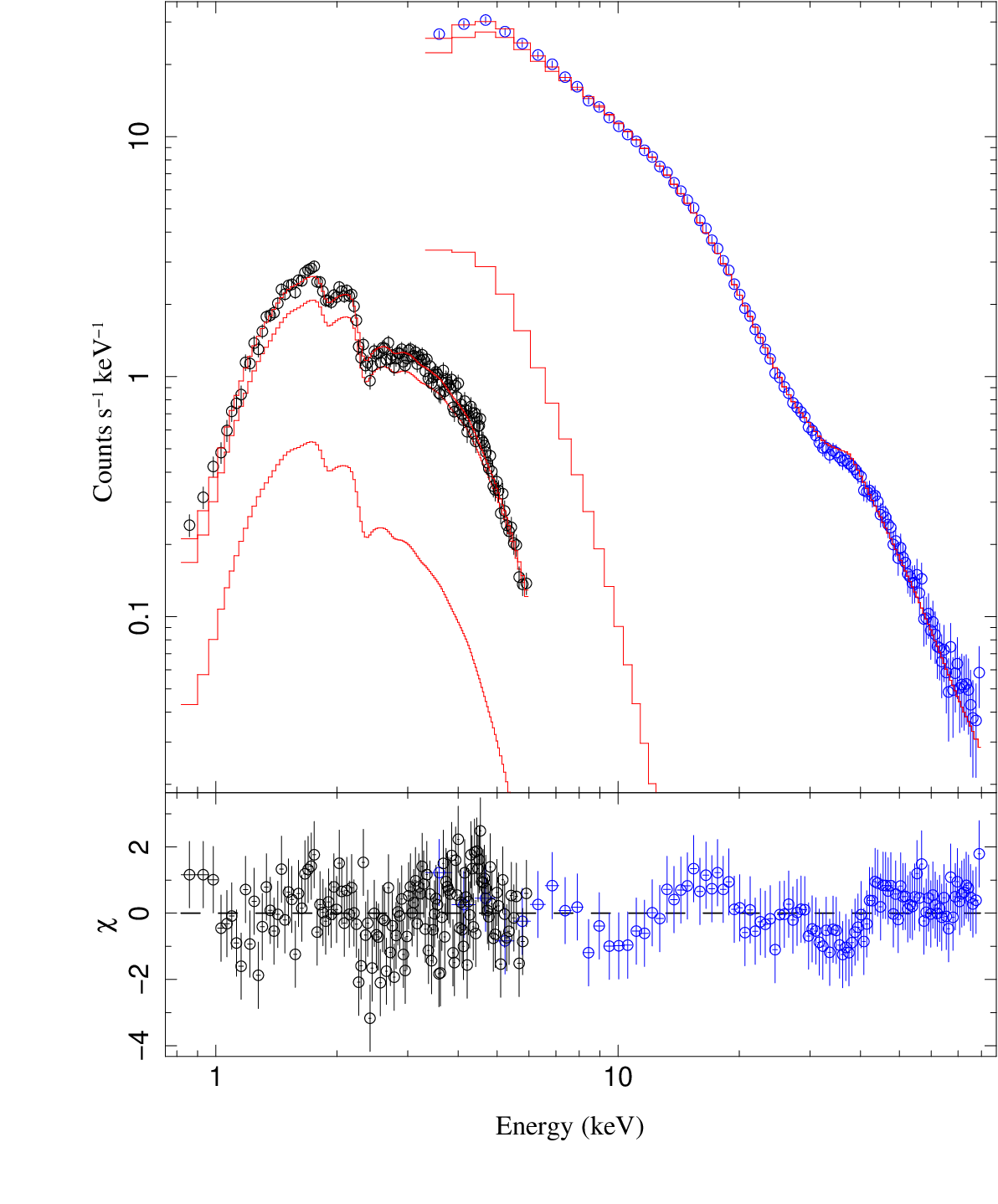}
\caption{The $0.7-80$ keV broadband X-ray energy continuum with the best fit model {\tt constant*tbabs(diskbb+nthcomp)}.
The black circles represent the SXT spectral data, whereas the blue ones are for the LAXPC spectral data.}\label{figOne}
\end{figure}


\begin{table*}[htb]
\tabularfont
\caption{Best-fit Temporal Parameters Obtained from the PDS Derived from LAXPC data}\label{tableExample} 
\begin{tabular}{lccccc}
\topline
Component & $3-15$ keV & $15-30$ keV & $30-80$ keV \\\midline
$\nu_{qpo}$ (Hz) & 0.596$\pm0.002$ & 0.598$\pm0.003$ & 0.597$\pm0.006$ \\ \\
Width$_{qpo}$ (Hz) & 0.075$^{+0.006}_{-0.005}$ & 0.071$\pm0.006$ & 0.07$\pm0.01$ \\ \\
Q$_{qpo}$ & 7.9$\pm0.6$ & 8.4$^{+0.7}_{-0.6}$ & 8.4 $^{+1.5}_{-1.3}$ \\ \\
rms$_{qpo}$ (\%) & 11.9$\pm0.3$ & 10.0$\pm0.3$ & 3.2$\pm0.2$ \\ \\
$\nu_{har}$ (Hz) & 1.20$\pm0.01$ & 1.22$^{+0.03}_{-0.02}$ & --- \\ \\
Width$_{har}$ (Hz) & 0.14$\pm0.03$ & 0.23$^{+0.12}_{-0.07}$ & --- \\ \\
Q$_{har}$ & 8.8$^{+2.1}_{-1.6}$ & 5.2$^{+2.5}_{-1.7}$ & --- \\ \\
rms$_{har}$ (\%) & 4.3$\pm0.4$ & 4.0$\pm0.6$ & --- \\ \\
$\nu_{bln1}$ (Hz) & 0.15$\pm0.01$ & 0.16$\pm0.01$ & --- \\ \\
Width$_{bln1}$ (Hz) & 0.44$^{+0.09}_{-0.07}$ & 0.27$^{+0.06}_{-0.05}$ & --- \\ \\
Q$_{bln1}$ & 0.35$^{+0.03}_{-0.04}$ & 0.60$\pm0.06$ & --- \\ \\
rms$_{bln1}$ (\%) & 9.3$\pm0.6$ & 6.1$\pm0.6$ & --- \\ \\
$\nu_{bln2}$ (Hz) & 1.08$^{+0.12}_{-0.14}$ & --- & --- \\ \\
Width$_{bln2}$ (Hz) & 1.9$\pm0.2$ & --- & --- \\ \\
Q$_{bln2}$ & 0.563$^{+0.008}_{-0.002}$ & --- & --- \\ \\
rms$_{bln2}$ (\%) & 11.8$^{+0.9}_{-1.1}$ & --- & --- \\ \\
$\nu_{bln(zero)}$ (Hz) & 0 (f) & 0 (f) & 0 (f) \\ \\
Width$_{bln(zero)}$ (Hz) & $>7.9$ & 3.5$^{+0.7}_{-0.5}$ & 0.81$^{+0.25}_{-0.20}$ \\ \\
rms$_{bln(zero)}$ (\%) & 9.7$^{+0.9}_{-0.7}$ & 11.9$^{+0.6}_{-0.7}$ & 3.5$\pm0.4$ \\ \\
$\chi^2$/dof & 215.7/126 & 163.6/129 & 174.7/135 \\
\hline
\end{tabular}
\tablenotes{f-indicates the fixed parameters.}
\end{table*}


\begin{table}[htb]
\tabularfont
\caption{Best fit spectral parameters for the joint fitting of SXT and LAXPC spectral data in $0.7-80$ keV band.}\label{tableExample} 
\begin{tabular}{lccccc}
\topline
Component & Parameter & Value \\\midline
TBabs & $N_H$ & 1.94$\pm0.07$ \\
Diskbb & KT$_{in}$ (keV) & 1.2$^{+0.1}_{-0.2}$ \\
		& Norm & 3.8$^{+1.4}_{-1.1}$ \\
NthComp & $\Gamma$ & 1.67$\pm0.02$ \\
		& KT$_e$ (keV) & $> 56.9$ \\
		& norm & 0.11$^{+0.02}_{-0.01}$ \\
		& $F_{total} (\times10^{-9})$ & 4.30$^{+0.06}_{-0.04}$ \\
		& $F_{disk}(\times10^{-10})$ & 1.3$\pm0.3$ \\
		& $\chi^2$/dof & 204.5/248 \\
\hline
\end{tabular}
\tablenotes{F- unabsorbed flux derived in the $0.7-80$ keV band.}
\end{table}

\vspace{2em}
\section{Discussion and Concluding Remarks}

In this paper, we have carried out a comprehensive spectral and temporal analysis of H~1743--322 with a single \textit{AstroSat} observation taken during the 2016 outburst of the source.
This observation was taken five days earlier than the simultaneous \textit{XMM-Newton} and \textit{NuSTAR} ones during the outburst rise phase of the source (see Figure 1).
 As similar to the study of the 2016 outburst of this source by Chand {\em et al.} (2020) using the above said simultaneous \textit{XMM-Newton} and \textit{NuSTAR} observations, QPO along with its upper harmonic in the commensurate ratio of $1:2$ are detected in both the $3-15$ keV and $15-30$ keV bands of the \textit{AstroSat} observation. For the first time, the LAXPC data onboard \textit{AstroSat} allows us to investigate the properties of PDS of H~1743--322 above 30 keV band. In the $30-80$ keV band, we also found the presence of QPO at the same frequency as in the above mentioned two bands. 
 The shape of the PDSs, as well as the values of the quality factor and the fractional rms amplitude of the QPOs in all the above mentioned three energy bands clearly indicate that the observed QPOs fall in the type C category (Casella {\em et al.} 2004, 2005; Motta {\em et al.} 2011). Moreover, the energy independent nature of the QPO and the upper harmonic indicates that the source was observed in the LHS during the outburst rise period with the luminosity in the X-ray regime capable enough to drive the transition of the source from the hard to soft state (Stiele \& Yu 2016). Similar kind of energy independent nature of type C QPOs along with upper harmonic was reported by Chand {\em et al.} (2020) for the same outburst. However, the centroid frequency of the QPO and its upper harmonic as found by Chand {\em et al.} (2020) with the simultaneous \textit{XMM-Newton} and \textit{NuSTAR} observations are shifted to higher frequency side with respect to this particular five days earlier \textit{AstroSat} observation by $\sim0.4$ Hz and $\sim0.8$ Hz, respectively. This indicates that the centroid frequency of the QPO and its upper harmonic were shifting to the higher frequency side as the time started to increase during the rising phase of the 2016 outburst. Similar kind of increase in the QPO frequency with the increase in time during the 1998  outburst rise phase was observed for the BHT XTE~J1550--564 (see Dutta \& Chakrabarti 2016). Since the QPOs are associated with the geometry of the corona, the shift in the centroid frequency of the QPOs may indicate certain geometrical change in the corona (see Titarchuk \& Fiorito 2004). It is found that the size of the corona decreases when the frequency of the QPO moves up during an outburst rise phase and the vice versa during a declining phase (Chakrabarti et al. 2008, 2009; Debnath et al 2010; Dutta \& Chakrabarti 2016). The overall decreasing trend of fractional rms amplitude of the QPO with energy clearly depicted in Figure 5 is similar to that found by Stiele \& Yu (2016) and Chand {\em et al.} (2020) for the 2014 and 2016 outbursts, respectively.

From broadband spectral analysis ($0.7-80$ keV) of the simultaneous SXT and LAXPC spectral data, it is found that the X-ray continuum can be described by the multi-colored disk blackbody model {\tt diskbb} modified by the galactic absorption {\tt tbabs}, and the thermal Comptonization model {\tt nthcomp}. The value of the column density (N$_H$) is found to be 1.94$\pm0.07$, which is consistent with those reported by previous workers using \textit{XMM-Newton}, \textit{NuSTAR}, \textit{Chandra}, \textit{Suzaku} and \textit{INTEGRAL} observations (Parmar {\em et al.} 2003; Corbel {\em et al.} 2006; Miller {\em et al.} 2006; Shidatsu {\em et al.} 2014; Stiele \& Yu 2016; Chand {\em et al.} 2020). The photon index ($\Gamma$) value of $\sim1.67$ indicates that the source was in the LHS during this particular \textit{AstroSat} observation.  The electron temperature ($KT_e$) in the coronal plasma is found to be $>56.9$ keV, which is quite higher relative to that reported by Stiele \& Yu (2016) during the 2014 outburst. The disk temperature (KT$_{in} \sim1.2$ keV) is also found to be higher compared to the 2014 outburst (see Stiele \& Yu 2016). Higher value of disk temperature is also reported by Chand {\em et al.} (2020) for the 2016 outburst, which may be due to the irradiation of the accretion disk by the high energetic coronal X-rays (Gierli{\' n}ski {\em et al.} 2009). We could not detect the presence of the iron line as reported by Chand {\em et al.} (2020) with \textit{XMM-Newton} and \textit{NuSTAR} observations, which may be due to the limitation of the SXT data up to 6 keV and large statistical uncertainties in the LAXPC data. It is clear from Table 2 that the disk fraction (unabsorbed (F$_{disk}/$F$_{total}$)) in $0.7-80$ keV is found to be $\sim3\%$, which is consistent with the value reported by Chand {\em et al.} (2020). Lower value of disk fraction indicates that the flux from the Comptonizing component has the maximum contribution to the total source flux.

As in Figure 6, a hard lag of $21.4\pm2.4$ ms was detected at the frequency range of $\sim1-5$ Hz between $3-10$ keV and $20-40$ keV bands. However, the lag amplitude is found to be weaker and detected at higher frequency as compared to that obtained by Chand {\em et al.} (2020) using the \textit{XMM-Newton} observations. Higher frequency lag observed in the \textit{AstroSat} data indicates the short term variability in the source. Morever, it is clear from Figure 2-4 and 6 that the QPO does not contribute significant power in the frequency range, where the time lag is detected. This indicates that the lag is not being triggered by the QPO rather it may arise due to the propagation of fluctuations in the mass accretion rate from outer part of the accretion disk to the inner hot regions (Lyubarskii 1997). This is also confirmed by the log-linear trend of the averaged time lag with energy (see Figure 7). Similar trend was observed with \textit{XMM-Newton} up to 10 keV by Chand {\em et al.} (2020) but this trend is extended to  higher energies up to 60 keV using \textit{AstroSat}. The absolute log-linear trend between the averaged time lag and energy may also rule out the possibility of hard lag, arising due to thermal Comptonization (Uttley {\em et al.} 2011, 2014; De Marco \& Ponti 2016). However, a soft X-ray lag was found during the 2008 and 2014 failed outbursts at the Eddington-scaled luminosity of $\sim0.004$ in the $3-10$ keV band, when the source was in hard state (De Marco {\em et al.} 2015; De Marco \& Ponti 2016). A similar value of the Eddington-scaled luminosity in the $3-10$ keV band was reported by Chand {\em et al.} (2020) using the \textit{XMM-Newton} and \textit{NuSTAR} observations taken during the 2016 successful outburst. Here, we also derived the Eddington-scaled luminosity for this particular \textit{AstroSat} observation by assuming the distance to the source and mass of the black hole as in De Marco \& Ponti (2016). The estimated Eddington-scaled luminosity appears to be $0.005\pm0.002$, which is similar to the values obtained by De Marco \& Ponti (2016) and Chand {\em et al.} (2020). This further confirms the fact that the change in the lag properties between the failed outbursts in 2008 and 2014 and the successful outburst in 2016 may be driven by the physical phenomena other than the source luminosity.  

\section*{Acknowledgements}

We thank the anonymous referee for useful comments that have improved the quality of the paper. The authors acknowledge the financial support of ISRO under \textit{AstroSat} archival Data utilization program (No: DS-2B-13013(2)/8/2019-Sec.2). This publication uses data from the \textit{AstroSat} mission of the Indian Space Research Organisation (ISRO), archived at the Indian Space Science Data Centre (ISSDC). This work has used the data from SXT and LAXPC instruments onboard \textit{AstroSat}. LAXPC data were processed by the Payload Operation Center (POC) at TIFR, Mumbai. This work has been performed utilizing the calibration data-bases and auxiliary analysis tools developed, maintained and distributed by the \textit{AstroSat}-SXT team with members from various institutions in India and abroad, and the SXT POC at the TIFR, Mumbai \url{(https://www.tifr.res.in/~astrosat_sxt/index.html)}. SXT data were processed and verified by the SXT POC. This research has also used the data from \textit{MAXI} space telescope provided by RIKEN, JAXA and the \textit{MAXI} team. VKA thanks GH,  SAG; DD, PDMSA and Director, URSC for encouragement and continuous support to carry out this research. P.T. expresses his sincere thanks to the Inter-University Centre for Astronomy and Astrophysics (IUCAA), Pune, India, for granting supports through the IUCAA associateship program. S.C. is also very much grateful to IUCAA, Pune, India, for providing support and local hospitality during his frequent visits.



\end{document}